# Video-rate high-precision time-frequency multiplexed 3D coherent ranging

Ruobing Qian[1], Kevin C. Zhou[1], Jingkai Zhang[1], Christian Viehland[1], Al-Hafeez Dhalla[1] & Joseph A. Izatt[1,2,*]

[1] Department of Biomedical Engineering, Duke University, Durham, NC, USA, 27708, USA

[2] Department of Ophthalmology, Duke University Medical Center, Durham, North Carolina 27710, USA

*Correspondence author: joseph.izatt@duke.edu

**Abstract:**

Frequency-modulated continuous wave (FMCW) light detection and ranging (LiDAR) is an emerging 3D ranging technology that offers higher sensitivity and resolution than conventional time-of-flight (ToF) ranging. Due to the limited bandwidth of digitizers and the speed limitations of beam steering using mechanical scanners, meter-scale FMCW LiDAR systems typically suffer from a low 3D frame rate (<1 Hz), which greatly restricts their applications in real-time imaging of dynamic scenes. In this work, we report a high-speed FMCW based 3D imaging system, combining a grating for beam steering with a compressed time-frequency analysis approach for depth retrieval. We thoroughly investigate the localization accuracy and precision of our system both theoretically and experimentally. Finally, we demonstrate 3D imaging results of multiple static and moving objects, including a flexing human hand. The demonstrated technique achieves submillimeter localization accuracy over a tens-of-centimeter imaging range with an overall depth voxel acquisition rate of 7.6 MHz, enabling densely sampled 3D imaging at video rate.

Real-time high-resolution three-dimensional (3D) imaging is highly desirable in a wide range of established and emerging fields including biomedical imaging, robotics, virtual/ augmented reality, 3D printing, and autonomous vehicles. A useful distinction can be made between 3D volumetric imaging systems which acquire fully sampled 3D tomographic data, versus 3D surface imaging or ranging systems which detect the depth range for every pixel in a 2D scene. The former are commonly used in medical imaging, whereas the latter, often collectively referred to as light detection and ranging (LiDAR), are of intense current interest primarily motivated by autonomous systems development. Optical coherence tomography (OCT) is a volumetric 3D imaging technique which has had great success in medical imaging, especially in ophthalmology [1,2]. To date, OCT has primarily been developed for applications requiring micrometer-scale resolution and millimeter-scale imaging range. Recently, swept-source OCT (SSOCT) systems using lasers sweeping at 100's of kHz to MHz repetition rate have enabled fully sampled 3D volumetric imaging with several millimeters imaging range at 10s of volumes/sec[3]. Frequency-modulated continuous wave (FMCW) LiDAR shares the same working principle as SSOCT, but prioritizes imaging range and high speed over axial resolution, and employs some form of surface detection (typically peak reflector position) to collapse the depth dimension to a single depth range measurement [4,5]. Compared to other 3D ranging technologies such as stereo cameras, structured light sensors and pulsed time-of-flight (ToF) LiDAR, FMCW LiDAR offers significant performance improvements in terms of higher detection sensitivity and immunity to ambient light.

One of the major challenges in FMCW LiDAR is extending the imaging range, which is determined by the instantaneous coherence length of the source and thus inversely related to the instantaneous linewidth. There is substantial current interest and effort in both the FMCW and OCT communities to increase the imaging range to the meter scale, of primary interest for room-scale robotic vision, and even multi-meter scale, which is of interest for autonomous vehicles. Multiple laser techniques have previously been demonstrated to decrease the instantaneous linewidth, such as vertical-cavity surface emitting lasers (VCSEL) [6-11], Fourier-domain mode locking(FDML) [12], optical time-stretching [13,14], single or stitched distributed feedback (DFB) [15,16] and akinetic all-semiconductor programmable lasers [17]. These techniques have enabled extensions of the OCT imaging range from several centimeters to 1.5m, and FMCW ranging with an imaging range up to several meters. However, increasing the imaging range in either technique without sacrificing depth resolution, using existing approaches, requires either a commensurate increase in receiver bandwidth or in acquisition time. Even with the fastest available photodetectors and digitizers, several seconds to hours have been required to acquire single volumes with hundreds of sampling points in each lateral dimension and tens of centimeters imaging range [8,11,15,17].

Another major challenge in FMCW LiDAR is the limited scanning speed, cost, and reliability of mechanical scanners. Traditional mechanical scanners, such as galvanometer mirrors and micro-electromechanical system (MEMS) scanners, have a maximum scanning rate of several kHz, which ultimately limits the line-scanning rate of both FMCW and ToF LiDAR. For these reasons, an emerging consensus in the LiDAR community is that compact, nonmechanical beam scanners with no moving parts are required for high speed imaging. For example, optical phased arrays (OPA) achieve beam steering by controlling the relative phase of arrays of coherent optical emitters [18-20], and thus allow compact solid-state LiDAR with MHz beam scanning bandwidths [21,22]. However, the beam steering performance of OPA degrades with substantial spatial sidelobes as the wavelength sweeps [23]. Another approach is to use diffractive optics to achieve spectrally encoded spatial scanning. For example, Jiang et.al, demonstrated a ToF LiDAR utilizing a discrete time-stretched broadband source and a grating for scanning [24]. Riemensberger et al., demonstrated a massively parallel FMCW LiDAR system using a frequency comb laser combined with a grating for fast-axis scanning and 30 parallel detection channels [4]. All these works require complex laser design and system architecture. Most recently, Okano et al., demonstrated a swept source FMCW ranging system using a tunable VCSEL laser and a diffraction grating for beam scanning [25]. However, this system still required several thousand spectral sampling points for each depth measurement, and thus only 30-45 distinct depth measurements were obtained along the grating scan axis. This limited the overall 3D voxel acquisition rate to 13.5-300 kHz, which is insufficient to support real-time densely sampled imaging operation.

Here, we report a time-frequency multiplexed FMCW LiDAR technique for high-speed high-precision 3D imaging using an akinetic all-semiconductor swept source, a diffraction grating for fast-axis beam steering, and a compressed sampling approach. In particular, we take advantage of the inherent sparsity of 3D ranging in the axial dimension, assuming that each lateral location contains only one dominant reflector in depth. Given this sparsity constraint, we show that we can reconstruct depth maps with higher precision and accuracy while using many fewer spectral points taken over a smaller bandwidth than the coherence length associated with that bandwidth. Using this optimized time-frequency analysis approach, we achieved sub-millimeter localization accuracy and precision using only 200 spectral points across a narrow bandwidth of 0.27nm for each depth measurement. A total of 475 depth measurements along the grating axis were obtained within a single laser sweep, generating an overall depth voxel acquisition rate of 7.6 MHz. We further studied the lateral and axial localization precision and accuracy of our system both theoretically and experimentally. Finally, we show real-time 3D imaging results of various everyday objects including a living human hand, with a maximum imaging range of 32.8cm and a 3D frame rate as high as 33.2Hz.

## Results

### Time-frequency Multiplexed FMCW LiDAR System Design

The schematic of the sample arm optical design is shown in Fig.1(a), along with zoomed-in views of the Zemax (Kirkland, WA) optical model in Fig.1(b-c). An akinetic all-semiconductor programmable swept laser source (Insight Photonics Solutions; Lafayette, CO) centered at 1316 nm was used (detailed in the Methods section). A collimated beam created by a reflective collimator was focused by a f=2m imaging lens. A galvanometer mirror was inserted before the imaging lens to accomplish beam scanning along the slow (vertical) axis, while a transmissive grating was placed immediately after the lens to achieve spectrally encoded scanning along the fast (horizontal) axis. The detailed optical design is described in the Methods section. The system achieved a horizontal FOV of 22.4 cm and diffraction limited performance with a lateral resolution of 890 µm at the focus (Fig.1(d-f)). The vertical FOV was determined purely by the scanning angle of the galvanometer mirror; a vertical FOV of 12-20cm was used in this work. The overall system design is shown in Fig.1(g). We employed a spectrally balanced interferometer topology incorporating three 50/50 2*2 fiber couplers (Thorlabs; Newton, NJ) [26], and the interferometric signal was detected using a balanced photodetector (Insight Photonics Solutions; Lafayette, CO) with 400 MHz bandwidth and digitized at 800MS/s (Alazar Technologies; Pointe-Claire, QC, Canada). A total of 47,646 samples was collected per sweep. The maximum imaging range of the system was approximately 32cm.

### Signal Processing

With a grating used for fast spectral scanning, the detected interferogram during a single sweep contains the signals from reflectors at different lateral positions along the horizontal axis. To retrieve the depth information of reflectors, instead of Fourier transforming the entire interferogram, a short-time Fourier transform (STFT) was applied (Fig. 2(a)). The spectral window size of the STFT determines the tradeoff between the angular/lateral resolution and axial resolution. A larger spectral window corresponds to a larger bandwidth, which leads to better axial resolution or localization accuracy, but fewer total number of windows, which leads to lower effective lateral resolution along the grating axis. Here, the optimal spectral window size was determined based on the lateral resolution of the system. As shown in Fig. 1(e), if point spread functions of two adjacent wavelengths are unresolvable (i.e. the centroid of the spot of one wavelength is within the Airy radius of the other wavelength), the detected signals from these two wavelengths were treated as arising effectively from the same position, and were therefore analyzed within the same spectral window. As such, we determined the optimal spectral window size of 1.56 $cm^{-1}$ in wavenumber or 0.27 nm in wavelength for our design, which corresponded to about 200 samples per window. The effective lateral resolution of the system along the grating axis was then determined by the sum of point-spread functions (PSFs) of all the wavelengths within the same spectral window, which is equivalent to the convolution of the PSF of single wavelength and the STFT window. Thus, the lateral resolution along the grating axis was determined to be 1240 µm in this design (see Supplementary Fig.1), while the lateral resolution along the galvanometer axis was still 890 µm, determined solely by the imaging optics as in conventional optical imaging.

The truncated signal from each spectral window (Fig.2(a)) was then zero-padded to 5000 samples before taking the Fourier transform (FT). Here, zero-padding enabled more accurate, sub-pixel peak localization. The depth of the dominant reflector was then localized if the peak intensity after taking the FT was above a predefined threshold (Fig.2(b)), otherwise that pixel was considered as background with no detected reflector and the depth value was assigned to not available (N/A). To achieve Nyquist lateral sampling, two adjacent spectral windows were overlapped by one half the window size (Fig.2(a)), which was 100 samples or 0.78 $cm^{-1}$ in wavenumber. Therefore, a total of 475 spectral windows were applied in each sweep/A-scan, which means 475 depth measurements (238 independent depth measurements) were obtained within a single sweep time of 62.7 µs. Finally, the depth map was acquired by scanning the beam in the slow axis using the galvanometer mirror, and performing the same STFT and peak localization analysis for each laser sweep.

**Localization Precision & Accuracy Characterization**

Our time-frequency multiplexed FMCW imaging technique is analogous to stochastic optical reconstruction microscopy (STORM) in the sense that both techniques are localization-based methods with corresponding sparsity requirements. STORM localizes the centroids of a subset of fluorophores that are activated at a given time with a typical precision down to tens of nanometers to achieve super-resolution imaging beyond the diffraction limit [27]. Similarly, our technique localizes the depth of an object assuming it has a dominant reflector (most likely the surface reflector) and achieves localization precision and accuracy better than the purely optical imaging resolution.

In the axial direction, we characterized both the axial localization accuracy and precision of the system. Here, axial localization precision refers to the uncertainty or repeatability of our axial localization, while axial localization accuracy refers to how close our depth measurement was to the ground truth depth.

**Same Spot Axial Localization Precision – Mirror and Diffuse Scattering Sample**

To quantify the axial localization precision for an ideal reflecting sample, we first imaged a gold mirror behind neutral density filters with various optical densities (ODs), which is a standard experiment in OCT to characterize axial resolution and sensitivity. We took data from 400 repeated laser sweeps on the mirror sample without any vertical galvanometer scanning, and calculated the SD of the retrieved depths at the same location along the horizontal grating scan direction (later referred to as the "same spot" axial localization precision). We performed these measurements at three different input power or SNR levels. Assuming the gold mirror is a perfect single reflector, the normalized interferogram S(k) can be modeled using the following equation,

$$S(k) = A\cos(2k\Delta z + \varphi) + Gauss\,(0, \sigma), \tag{1}$$

where A and $\varphi$ are amplitude and phase of the interference fringe, and $\Delta z$ is the depth of the reflector. Assuming the system is shot noise limited, in the limit of a large number of photons the noise is approximately normally distributed, $Gauss\,(0, \sigma)$, with an SD of $\sigma$. Based on this model, the minimum theoretical localization uncertainty of $\Delta z$ has previously been derived [28], which in practice can be achieved using the aforementioned Fourier-domain zero-padding approach[28]. The SD of the $\Delta z$ localization, $\delta z$, can be estimated using the following equation[28],

$$\delta z \approx \left(1.6 \times SNR \times 2\Delta k\sqrt{N_s}\right)^{-1}, \tag{2}$$

where $\Delta k$ is the total bandwidth in wavenumber, and $N_s$ is the total number of spectral sampling points. In our system, the number of sampling points, $N_s$, in each STFT window was 200, and the bandwidth of each window, $\Delta k$, was 1.56 cm$^{-1}$. Here, the SNR of the detected interferogram is defined as[28],

$$SNR = \frac{A}{\sqrt{2}\sigma}. \qquad (3)$$

Using this equation, we calculated the SNRs of our detected interferograms, and a plot of the SDs of the retrieved depths at three different SNR levels is shown in the dark blue line in Fig. 3(a). Assuming 100% sweeping linearity, the theoretical SDs of depth localization at the same SNR levels as the experimental data are calculated using Eq. 2 and plotted in red line in Fig. 3(a). We then compared our experimental results with the theoretical predicted axial localization precisions at these corresponding SNR levels. As expected, the localization precision increased as the SNR of the detected signal increased (experiment: from 46.5 µm to 41.6 µm; simulation: 9.69 µm to 3.79 µm). However, the measured SDs from our experiment were more than 4x larger than the theoretical SDs.

The model of Eq.2 assumes 100% laser sweeping linearity, such that adjacent spectral sampling points are evenly spaced in wavenumber. To also include the effect of sweeping nonlinearity, we extend the model to include sweep nonlinearity using the following equation

$$S(k) = A\cos(2(k + Gauss(0, \sigma_k))\Delta z + \varphi) + Gauss(0, \sigma). \qquad (4)$$

Here, $Gauss(0, \sigma_k)$ is the gaussian wavenumber nonlinearity noise with a SD of $\sigma_k$. Using this extension to the model, we simulated S(k) with $\Delta z$ the same as the experiment data, and various $\sigma_k$ ranging from 0.1 pm (5.8*10$^{-4}$ cm$^{-1}$) to 1 pm (5.8*10$^{-3}$cm$^{-1}$) at the same SNR levels and calculated the SD of depth localization. The results are plotted in green, black and cyan lines in Fig. 3(a). Our experiment results are closest to the simulation results with ±0.5pm nonlinearity (black line), which is also the nonlinearity specification provided by the laser manufacturer [29]. Thus, in our system, it is clear that the sweep nonlinearity played a more significant role in determining the same spot axial localization precision in this application.

On the other hand, the OCT axial resolution, defined as the coherence length of the laser over the wavelength sweep range per acquisition, is calculated using the below equation,

$$l_c = \frac{2 \ln 2}{\pi} \frac{\lambda_0^2}{\Delta \lambda}, \qquad (5)$$

where $\lambda_0$ and $\Delta\lambda$ are the central wavelength and the bandwidth of the source [30]. In our system, the bandwidth of each STFT window was 0.27 nm, and the corresponding OCT axial resolution (i.e., coherence length associated with that bandwidth) was 2.82 mm. Meanwhile, our measured localization precisions were 41.6-46.5 µm, indicating that our same spot localization precision for an ideal mirror sample was >60x better than the theoretical axial resolution.

We note that for a grating-scanned system with non-telecentric scanning, a mirror or other specularly dominant reflector is useful for comparison with theory, but the results for localization accuracy and precision may differ from real-world diffusely scattering samples since a mirror artificially enforces an exact back-scattering requirement. Therefore, for the remainder of our localization accuracy and precision measurements, in order to depict real-world performance, diffuse scattering samples were used.

For our repeated same spot measurement of the diffusely scattering anodized aluminum sample, the SD of depth localization at a single lateral position on the sample was 64.2 µm. While we did not have means to accurately measure the SNR of the signal from this sample, this result was not significantly worse than the mirror results for which the SNRs were known as depicted in Fig. 3(a). (41.6-46.5 µm).

**Scanning Spot Axial Localization Precision & Accuracy – Diffuse Scattering Sample**

To characterize any additional contributions to axial localization uncertainty arising from lateral scanning, we performed "scanning spot" measurements using two staggered metal base plates separated by 25.4mm in depth (Fig. 2(c), detailed in the Methods section). Our measurements results characterizing the scanning spot axial localization precision and accuracy for the anodized aluminum sample are depicted in Fig. 3(b-c), in which the axial localization accuracy results measured at five different depths (~4, 10, 16, 22 and 28 cm) using the staggered metal plate sample are shown. The SDs of depth localization obtained from 100 different positions at the front surface and the back surface along two lines (Fig. 2(c)) are plotted in Fig. 3(b), and the mean measured depth differences at five different depths were 25.30 mm, 25.44 mm, 25.27mm, 25.48mm and 25.14mm, which were all close to the ground truth, 25.40 mm (Fig. 3(c)). Interestingly, the SDs of the scanning spot depth localization measurements across the metal sample (~500-800 µm) were ~7-11x larger than the SD of the same spot depth localization at a single lateral position in the same metal sample (64.2 µm). Since the surface roughness of these professionally machined surfaces is expected to be substantially less than 0.5mm, this difference is likely due to independent realizations of speckle as a function of lateral position arising from the distribution of sub-resolution reflectors in the diffusely scattering sample, including those below the surface. Nevertheless, even for this real-world sample, the measured axial localization precision still exceeded the theoretical OCT axial resolution, given the laser bandwidth used to obtain the depth measurement, by ~4x. We note that scanning spot precision and accuracy are clearly sample dependent. The values are expected to be closer to the same spot precision when the sample is more like a specular reflector (i.e. a dominant single reflector with minimal surface roughness).

**Lateral Localization Precision – Diffuse Scattering Sample**

Finally, we characterized the lateral localization precision of our system by imaging the same staggered metal plates and quantifying the uncertainty of the edge localization (detailed in the Methods section). The horizontal and vertical localization precisions measured using the same staggered metal piece at five depths (~~4, 10, 16, 22 and 28 cm) and three different lateral positions (center: ~[0,0]cm; edge: ~[+10,0]cm; corner: [+10, +10]cm) are shown in Fig. 3(d). Overall, the localization precisions in both directions were uniform across the imaging depth and lateral FOV. The mean and SD of vertical localization precision was 143.2 µm and 24.8 µm, while the mean and SD of the horizontal localization precision was 205.4 µm and 26.7 µm. It was expected that the vertical localization precision would be better than the horizontal localization precision, as the vertical resolution (~890 µm) of our imaging system is better than the horizontal resolution (~1240 µm) due to STFT analysis. Nevertheless, this result demonstrates that our system localized the reflector depth variation laterally better than the optical resolution of the system.

**3D Imaging Results on Everyday Samples**

We performed 3D imaging on multiple static samples and a living human hand. First, to demonstrate the long axial imaging range of our system, we imaged two ceramic coffee cups (Fig.4(d)) which were axially separated by >9cm. 1000 scans across a vertical FOV of 15cm were acquired, which corresponds to a 3D imaging frame rate of 15.94Hz. The processed depth map with 475*1000 pixels (spanning 22.3 x 15cm) and the corresponding 3D rendering of cups are shown in Fig.4(a-b). In Fig.4 (c), we plot the cross-section depth profile along the black line in Fig.4(a)); the contour of the cup can be clearly observed.

To demonstrate that this technology can image objects with relatively weak surface reflections, we first imaged a synthetic rubber mannequin head (Fig.4(h)). Similar to the coffee cup imaging, 1000 scans across a vertical FOV of 15cm were acquired, corresponding to the same 3D imaging frame rate of

15.94Hz. Due to relatively weak scattering signal from the sample, the intensity thresholding-based depth localization approach was not sufficient to localize the depth of every reflector within the sample. Additionally, when a lower threshold value was applied, more background stripe noise was introduced due to imperfect removal of invalid points during the transitions between subintervals of a laser sweep [31] (Supplement Fig. 2(a)). Therefore, we applied a gradient-based background noise removal algorithm along with a 3*3 median filter to create the final depth map and 3D volume rendering (see Supplement Fig.2). The representative depth map and the corresponding volume rendering of the head are shown in Fig. 4(e-f). In Fig. 4 (g), we plot the cross-section depth profile along the black line in Fig. 4(e). The contours of the forehead, nose, upper and lower lip can clearly be resolved.

Finally, we demonstrate that our technology can not only achieve video-rate 3D ranging that allows us to image moving objects, but also be applied for in vivo imaging. Here, we imaged a hand adjacent to a metal stage and actively making a fist. 480 scans (including 80 scans for galvanometer flyback) across a vertical FOV of 16cm were acquired, which corresponds to a frame rate of 33.2Hz. The final depth maps with 475*400 pixels (spanning 22.3 x 16cm) and the corresponding 3D renderings of the hand at different times are shown in Fig.4(i-j). For the depth maps in Fig.4(i), the same noise removal method discussed for the face in 4(e-f) was applied. For the volume rendering in Fig.4(j), to further remove the background stripe noise, an 8*8 median filter was used. A ~2s video (67 frames) of the volume rendering played in 0.5x speed is shown in Supplement Movie 1. Although human skin is a relatively weakly scattering sample, the depth map of skin is still retrieved with high axial localization accuracy.

## Discussion

Our time-frequency multiplexed FMCW system achieves high-speed high-precision 3D imaging by using a broadband swept source with narrow instantaneous linewidth and a diffraction grating for spectrally encoded fast axis scanning. By applying a compressed sampling approach using an optimized window size and zero padding, 238 independent depth measurements along the grating axis were obtained within a single sweep time. Although each window had a narrow bandwidth of 0.27nm, we demonstrated on both mirror and metal samples that the axial localization accuracy and precision were significantly better the theoretical resolution, which was nearly 3mm. 3D imaging of multiple static samples and video-rate imaging of a moving human hand demonstrate the great potential of this technology in a wide range of potential applications in the fields of robotics navigation, virtual reality and 3D printing.

One major limitation of our current imaging system is the relatively short imaging depth range of 32cm, which is currently limited by the bandwidth of our available photodetector and digitizer. Using a commercially available higher speed digitizer, 4G/s, along with a similarly available photodetector with sufficient bandwidth (>2GHz), the imaging range of our system could be further extended to more than 1.6m. An increase of sampling rate would also lead to more sampling points per STFT window, improving the localization precision as predicted by Eq. (3). To further increase the imaging range, we can reduce sweep rate of the laser with a tradeoff of frame rate.

Another limitation of our current system is the limited lateral FOV along the horizontal axis. The horizontal angular FOV is fundamentally determined by the bandwidth of the source and the groove density of the grating. With a 65nm bandwidth centered at 1316nm and a 1145 grooves/mm grating, an angular FOV of 7.1º was achieved. To increase the angular FOV, a source with a larger bandwidth or a grating with a larger groove density could be used. Or, without changing the angular FOV, the lateral FOV could also be extended by simply increasing the working distance, although an imaging lens with a longer focal length or even a collimated beam needs to be used to extend the axial location of the focal plane. A telescope could also be added after the grating to expand the angular FOV.

A total of 238 independent depth measurements were obtained within a single sweep with current setup. To increase the number of independent measurements per sweep without changing the laser and grating, a narrower STFT window needs to be used, which means a more tightly focused beam with better lateral resolution is needed. However, this will potentially lead to worse axial localization accuracy and precision, as explained in Eq. (3), as well as shorter depth of focus. Therefore, the numerical aperture (NA) of the imaging beam determines the tradeoff between the total number of resolvable measurements along the grating axis and the axial localization precision and depth of focus. The optimal choice of the NA depends on the applications and will be further investigated in the future studies.

Our measured same spot axial localization precision on both mirror and machined metal samples was about an order of magnitude better than the measured scanning spot axial localization precision on the metal sample. Here, the same spot axial localization precision can be considered as the system-limited axial localization precision, since this value only depends on the SNR of the detected signal and sweep linearity of the source as shown above, while the scanning spot localization precision can be considered as the sample-limited axial localization precision, as it includes any additional contributions to axial localization uncertainty arising from sample roughness or other deviations from the single reflector assumptions. It is notable that the sample-limited axial localization precision on the mirror sample could not be measured, since the optical system was not telecentric and thus only a very small central region of the mirror satisfied the exact backscattering requirement. The difference between system-limited and sample-limited localization precisions on the metal sample is likely due to the effects of speckle. Since the adjacent wavelengths in each STFT window are not completely overlapped, and the metal is not a single specular reflector, different wavelengths interact with different subresolution reflectors, leading to speckle. With a narrow bandwidth of 0.27nm, the effect of speckle is even more significant, as also observed in a previous grating-based scanning microscopy system[32]. To reduce this effect, besides using a broader bandwidth source and a larger STFT window, methods such as spatial or angular compounding could also be considered[33,34].

In conclusion, we demonstrated a time-frequency multiplexed FMCW ranging system that combines grating-based spectrally encoded fast scanning and optimal STFT analysis for depth retrieval. The system can perform video-rate high-precision 3D imaging with an imaging range of 32cm, and potentially be used in many emerging industrial, automotive and biomedical fields.

## Methods

**Akinetic All-semiconductor Programmable Swept Source**

An akinetic all-semiconductor programmable swept laser source (Insight Photonics Solutions; Lafayette, CO) centered at 1316 nm with a 65.85 nm bandwidth was used. The source has an output power of ~70 mW and a nearly flat power spectrum across the entire bandwidth. The source also has an instantaneous coherence length of >1m and a linearity $\leq \pm 0.5$pm root mean square without the need for an external k-clock [29]. The sweep rate of the laser can be adjusted from 10 kHz to 200 kHz, and for this application, we set the sweep rate of the laser at 15.94 kHz.

**Time-frequency Multiplexed FMCW LiDAR Optical Design**

A 4mm diameter collimated beam was created by a reflective collimator (Thorlabs; Newton, NJ), and then focused by a 2m focal length plano-convex lens (Thorlabs; Newton, NJ) (Fig.1(a)). A galvanometer mirror (Thorlabs; Newton, NJ) was inserted before the imaging lens to accomplish beam scanning along the slow (vertical) axis, while a 1145 grooves/mm volume phase holographic transmissive grating (Wasatch Photonics; Logan, UT) was placed immediately after the lens to achieve spectrally encoded scanning along the fast (horizontal) axis. The grating was positioned with an incident angle of ~48° to maximize diffraction efficiency, and the input beam power after the grating was 15.2 mW. With the total sweep bandwidth of the input source of 65.85 nm, the angular FOV along the horizontal axis was 7.1°. With a working distance (from the grating to the focal plane) of 196 cm, the system thus achieved a horizontal FOV of 22.4 cm and diffraction limited performance with a lateral resolution of 890 μm at the focus (Fig.1(d-f)).

**Scanning Spot Axial Localization Precision & Accuracy**

To characterize scanning spot axial localization precision and accuracy, we imaged two staggered metal base plates separated by 25.4mm in depth (Fig. 2(c)). We measured the depth profiles (Fig. 2(d)) along two lines in the galvanometer scan direction (blue and black lines in Fig. 2(c)) of the front and back metal plates. We defined the scanning spot axial localization precision as the SD of the depth localization from multiple laterally displaced locations along the galvanometer scan direction line, and the scanning spot axial localization accuracy as their depth difference (Fig. 2(e), to be compared to the ground truth depth difference of 25.4 mm (red line in Fig. 2(e). We repeated this measurement at five different axial positions (~4, 10, 16, 22 and 28 cm).

**Lateral Localization Precision**

We characterized the lateral localization precision of our system by imaging the same staggered metal plates and quantifying the uncertainty of the edge localization. To measure the horizontal localization precision, we aligned the edge of metal plates perpendicular to the grating/horizontal axis, and obtained the depth map using the same STFT processing method described above, except that for characterization purposes, we oversampled the depth map by having two adjacent STFT windows separated by 10 samples or 0.08cm$^{-1}$ in wavenumber. We plotted the resulting edge response function (Fig. 2(f)) along the horizontal axis (red line in Fig. 2(c)) and obtained the depth range measurement as the peak of its derivative (Fig. 2(g)). We repeated this measurement at 250 consecutive lateral positions along the vertical axis. Finally, we fit the peak location profile with a line to remove the residual tilt due to imperfect alignment of metal plates, and calculated the residual standard deviation (Fig. 2(h)), to arrive at the horizontal localization precision. Using the same approach, we measured the vertical localization precision by imaging the same metal base plates rotated by 90 degrees. We placed the base plates at three different lateral positions(center: ~[0,0]cm; edge: ~[+10,0]cm; corner: [+10,+10]cm) and five different axial positions (~4, 10, 16, 22 and 28 cm) to quantify the horizontal and vertical lateral localization precision over the entire 3D imaging space.

## Data availability

All relevant data are available from the authors upon request.

## Code availability

All relevant codes are available from the authors upon request.

# Acknowledgments

This work was supported in part by NIH EY025009, NSF CBET-1902904, and DOD W81XWH-16-1-0498. The authors would like to thank Michael Crawford from Insight Photonics Solution for his assistance in laser setup.

## Author contributions

R.Q., K.C.Z., and J.A.I. conceived and developed the idea. R.Q. designed and built the system, analyzed the data. R.Q. and J. Z. collected the data. K.C.Z developed the theoretical model. R.Q. and A.D. performed the Zemax simulations. C.V. developed the data acquisition software. J.A.I. supervised the work. R.Q. wrote the manuscript with input from all authors

## Competing interests

Duke University has a patent pending on the technology described in the manuscript.

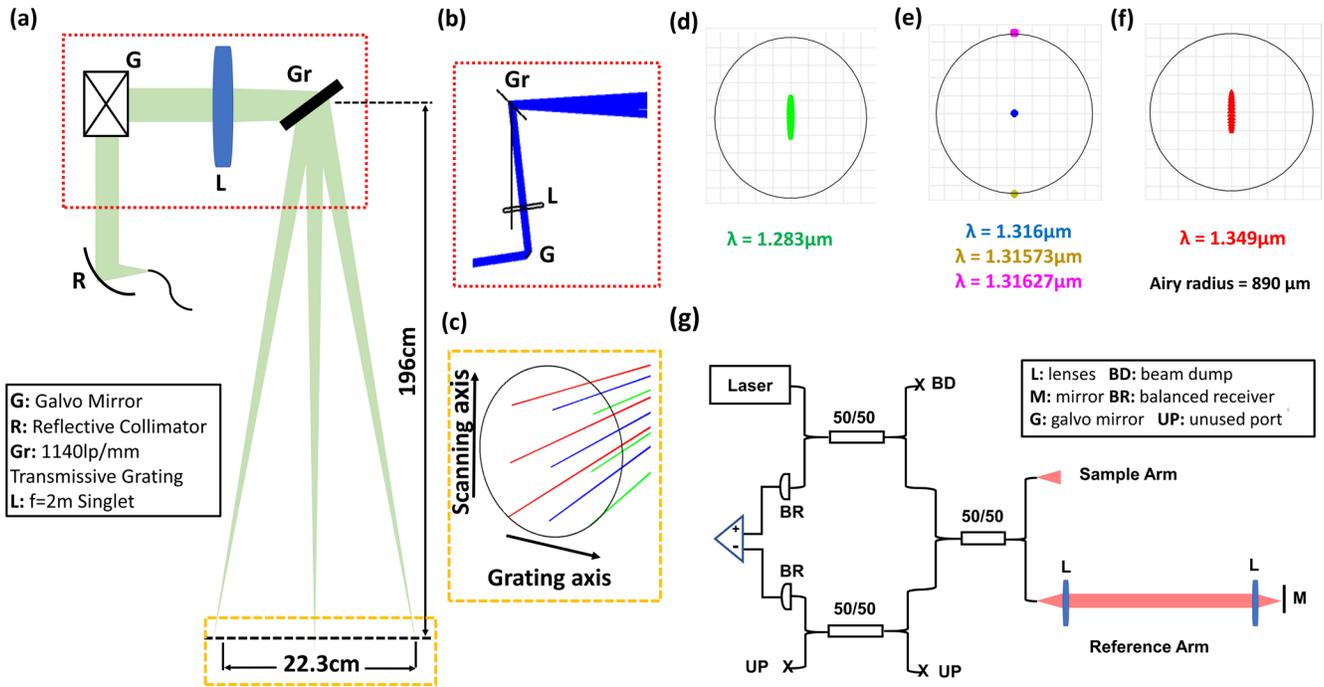

**Figure 1: Optical and system design of the real-time time-frequency multiplexed FMCW LiDAR system. (a)** Schematic of the sample arm optical design and **(b-c)** corresponding zoomed-in Zemax design; **(d-f)** Spot diagrams at the imaging plane of the **(d)** first, **(e)** central and **(f)** last wavelengths, along with **(e)** two wavelengths that have centroids of spots located adjacent to the central wavelength with a separation distance of the Airy radius at the focus; **(g)** Schematic of the overall system design.

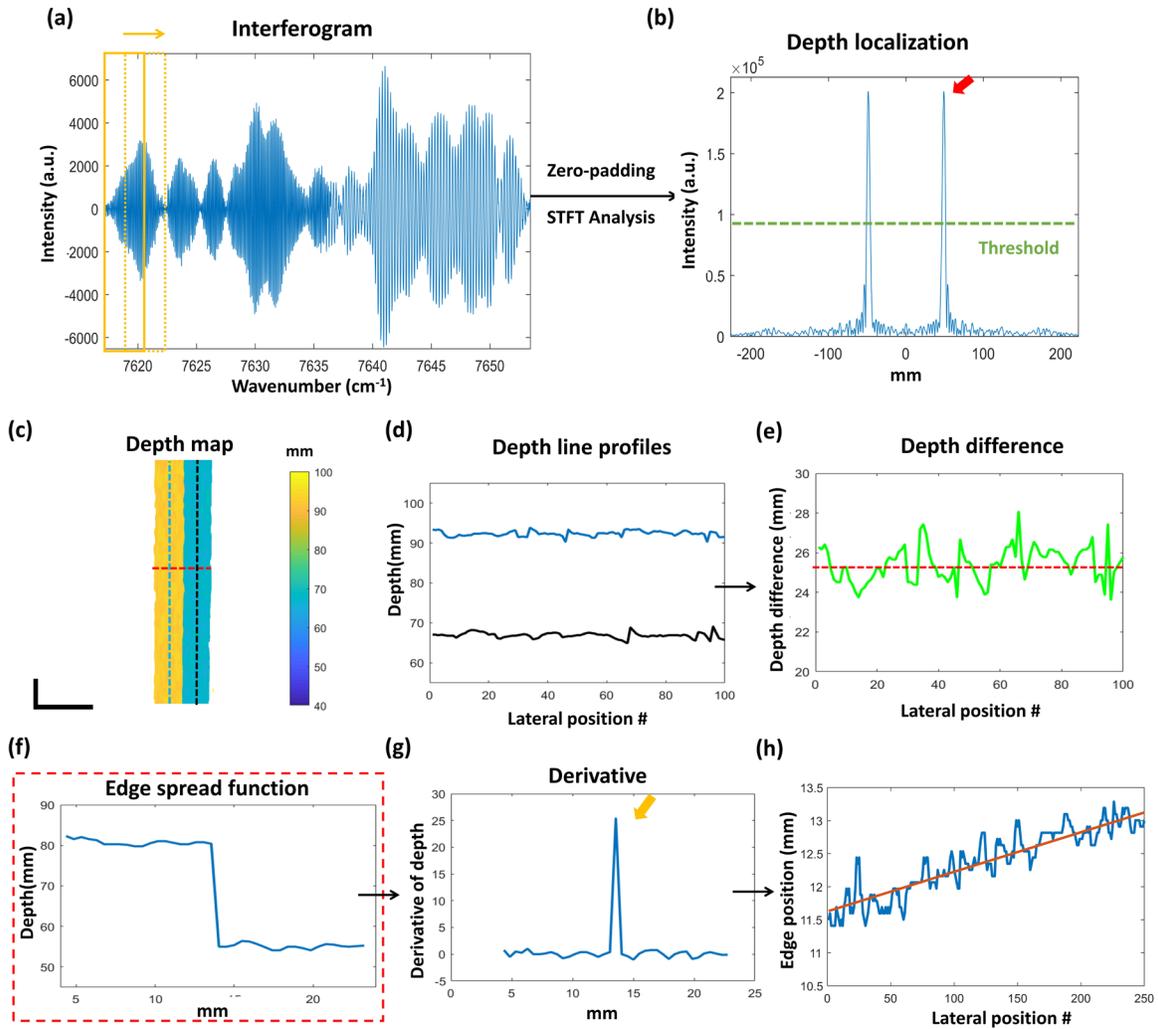

**Figure 2: Signal processing and system performance characterization of the time-frequency multiplexed FMCW LiDAR system. (a)** Short-time Fourier Transform (STFT) analysis to retrieve the depths of reflectors along the grating axis and **(b)** depth localization from the signal after FT and. **(c)** Depth map of two staggered metal base plates separated by 25.4mm, scale bar: 2cm. **(d-e)** The "scanning spot" axial localization precision and accuracy characterization: **(d)** depth profiles along two lines of the front (black line in **(c)**) and back metal plates (blue line in **(c)**) and **(e)** their depth differences. **(f-h)** The lateral localization precision characterization: **(f)** the edge spread function of the depth profile along the red line in **(c)** and **(g)** the edge localization from its calculated derivative function, and **(h)** the residual localization error measured at 250 lateral positions.

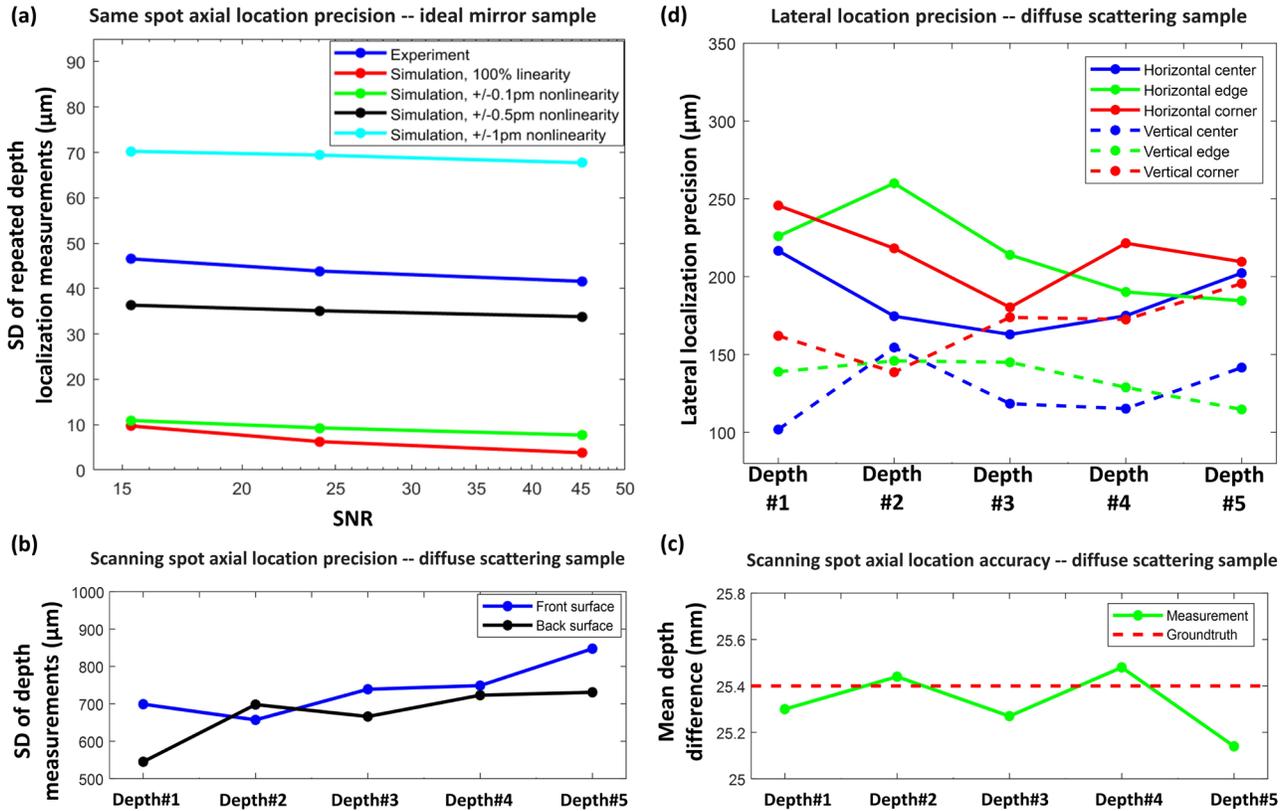

**Figure 3: Results of system localization precision and accuracy characterization. (a)** Measured axial localization precision calculated from the standard deviation (SD) of repeated depth measurements at the "same spot" in the grating scan direction, using an ideal mirror sample. Also plotted are the theoretical axial localization precisions with various wavenumber nonlinearity coefficients (0, ±0.1, ±0.5 and ±1 pm) at three different signal-to-noise ratios (SNRs). Measured axial localization **(b)** precision and **(c)** accuracy calculated from the SD of depth measurements from multiple "scanning spots" along two lines in the galvanometer scan direction using a diffuse scattering sample (two staggered anodized aluminum plates). Plotted are **(b)** the SD of the depths of the front and back surfaces of the plate and **(c)** the mean depth difference at five different depths (~4, 10, 16, 22 and 28 cm). **(d)** Horizontal and vertical localization precisions at three different depth locations (~4, 10, 16, 22 and 28 cm) and three different lateral positions (center: ~[0,0] cm; edge: ~[10,0] cm; corner: [10,10] cm) within the measurement field-of-view.

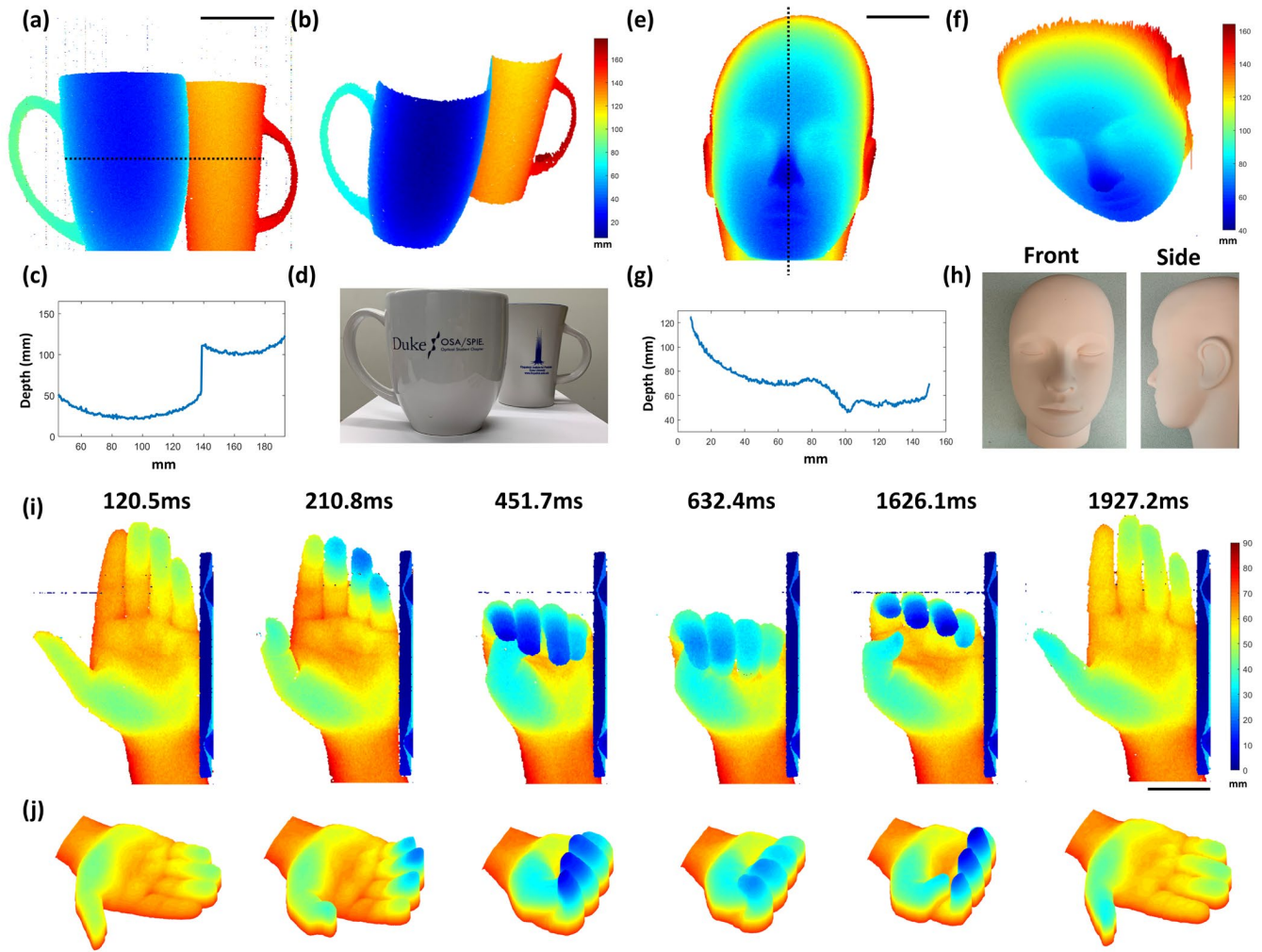

**Figure 4: Video-rate 3D imaging results of real-world objects. (a-d)** Imaging results of two ceramic coffee cups axially separated by >9cm, imaged with 475x1000 depth pixels at 16Hz : **(a)** depth map and **(b)** 3D surface rendering of the cups; **(c)** cross-section depth profile along the black line in **(a)**; (d) photo of the cups. **(e-h)** Imaging results of a mannequin head, imaged with 475x1000 depth pixels at 16Hz: **(e)** depth map and **(f)** 3D surface rendering of the mannequin head; **(g)** cross-section depth profile along the black line in (e); **(h)** Front and side views of the mannequin head. **(i-j)** Imaging results of a human hand adjacent to a metal stage and actively making a fist, imaged with 475x400 depth pixels at 33.2Hz: **(i)** depth maps and **(j)** 3D surface renderings of the hand while flexing; shown are 6 individual frames from a 2s video acquired with a frame rate of 33.2Hz. See Supplement Movie 1 for the complete video. Scale bar: 5cm.